\documentstyle[12pt]{article}

\makeatletter                                                           

\def\section{\@startsection {section}{1}{\z@}{-1.5ex plus -.5ex         
minus -.2ex}{1ex plus .2ex}{\large\bf}}                                 

\def\@thmcountersep{}                                                   

\long\def\@makecaption#1#2{\vskip 10pt \setbox\@tempboxa\hbox{#1. #2}   
   \ifdim \wd\@tempboxa >\hsize   
       #1. #2\par                 
     \else                        
       \hbox to\hsize{\hfil\box\@tempboxa\hfil}                         
   \fi}                                                                 

\def\ps@headings{                                                       
 \def\@oddhead{\footnotesize\rm\hfill\runninghead\hfill}               
 \def\@evenhead{\@oddhead}                                              
 \def\@oddfoot{\rm\hfill\thepage\hfill}\def\@evenfoot{\@oddfoot} }      

\makeatother                                                            

\title{The First Law of Black Hole Mechanics}{}{}

\def\runninghead{WALD :\quad FIRST LAW}

\author{
{\em Robert M. Wald} \thanks{Enrico Fermi Institute and Department
of Physics,
 University of Chicago, 5640 S. Ellis Avenue, Chicago, IL 60637, USA.
This research was supported in part by the National Science Foundation
under Grant No. PHY89-18388.}
}

\date{} 

\begin{document}

\pagestyle{headings}                                                   
\flushbottom                                                           

\maketitle
\vspace{-10pt} 

\begin{abstract}

A simple proof of a strengthened form of the first law of black
hole mechanics is presented.  The proof is based directly upon the
Hamiltonian formulation of general relativity, and it shows that the
the first law variational formula holds for arbitrary
nonsingular, asymptotically flat perturbations of a stationary,
axisymmetric black hole, not merely for perturbations to other stationary,
axisymmetric black holes.  As an application of this strengthened form of
the first law, we prove that there cannot exist Einstein-Maxwell
black holes whose ergoregion is disjoint from the horizon.  This
closes a gap in the black hole uniqueness theorems.

\end{abstract}

\section{Derivation of the First Law}
\indent
It was noted by Hilbert at the inception of general relativity that the
Einstein field equations are derivable from an action principle,
\begin{equation}
S = \frac{1}{16\pi} \int R \sqrt{-g}\; d^4x
\end{equation}
Thus, general relativity has a Lagrangian formulation. The corresponding
Hamiltonian formulation was given many years later in a collaboration
between Charles Misner, Richard Arnowitt, and Stanley Deser.  The main
results of this collaboration are summarized in [1].

        The Hamiltonian formulation of general relativity is employed as a
starting point in all attempts to formulate a quantum theory of gravity via
the canonical approach.  It plays a less essential role within the context of
purely classical general
relativity.
However, even in that context, the Hamiltonian formulation of general
relativity provides some penetrating insights into the structure of the theory.
In this paper, I shall illustrate this point by showing how a strengthened
form of the first law of black hole mechanics can be derived in  a
very simple and direct manner from the Hamiltonian formulation of
general relativity.  The results presented here were obtained in
collaboration with D. Sudarsky and were first reported in [2].  I
shall restrict attention here to Einstein-Maxwell theory--the more
general case of Einstein-Yang-Mills theory was considered in [2]--but
the analysis generalizes straightforwardly to allow other fields,
provided only that a Hamiltonian formulation of the complete theory
can be given.

        In the Hamiltonian formulation of Einstein-Maxwell theory, a point
in phase space corresponds to the specification of the fields
($h_{ab},\pi^{ab}, A_a,
E^a$) on a three dimensional manifold
$\Sigma$.
Here $h_{ab}$ is a Riemannian metric on $\Sigma$ and $A_a$  is
 the spatial part of the vector potential (i.e., the pull-back to
$\Sigma$   of the spacetime vector potential $A_{\mu}$).
The momentum canonically conjugate to $ h_{ab}$ is $\frac{1}{16
\pi}\;\; \pi^{ab}$,
where $\pi^{ab}$ is related to the extrinsic curvature, $K_{ab}$, of
$\Sigma$ in
the spacetime obtained by evolving this initial data by,
\begin{equation}
\pi^{ab} = \sqrt{h} (K^{ab} - h^{ab} K)
\end{equation}
The momentum conjugate to $A_a$  is $\frac{1}{4 \pi} \; \sqrt{h}
\; E^a$, where $E^a$  is the electric field in the evolved spacetime.

        Constraints are present in Einstein-Maxwell theory. The allowed
initial data is restricted to the constraint submanifold in phase space
defined by the vanishing at each point $x \epsilon \Sigma$
of the following quantities,
\begin{equation}
0 = {\cal C} = \frac{1}{4 \pi} \sqrt{h} D_aE^a
\end{equation}
\begin{equation}
0 = {\cal C}_0 = \frac{1}{16\pi} \sqrt{h}\; \{ -R + 2E_aE^a+F_{ab}F^{ab} +
\frac{1}{h}
 (\pi^{ab}\pi_{ab} - \frac{1}{2} \pi^2) \}
\end{equation}
\begin{equation}
0 = {\cal C}_a = - \frac{1}{8 \pi} \sqrt{h}\; \{D_b(\pi_a^b/\sqrt{h}) - 2
F_{ab} E^b \}
\end{equation}    
where $D_a$ is the derivative operator on $\Sigma$ compatible with
$h_{ab}$, $R$  denotes the scalar curvature of $h_{ab}$, and
$F_{ab} = 2D_{[a}A_{b]}$.

The ADM Hamiltonian, $H$,  for Einstein-Maxwell theory has the
``pure constraint'' form,
\begin{equation}
H = \int_{\Sigma} (N^\mu {\cal C}_\mu + N^\mu A_\mu {\cal C})
\end{equation}  
Here $N^\mu$  and $A_0$  are to be viewed as non-dynamical variables,
which may be prescribed arbitrarily. In the spacetime obtained by solving
Hamilton's equations, $N^\mu$  has the interpretation of being the time
evolution vector field (i.e., its projection
normal to $\Sigma$ yields the lapse function, $N$, and its projection
into $\Sigma$ yields the shift vector, $N^a$), and $A_0$ has the interpretation
of being the component of the vector potential normal to $\Sigma$.
The ``pure constraint'' form of $H$ is not special to Einstein-Maxwell
theory; any Hamiltonian arising from a diffeomorphism invariant
theory always takes such a form (see the appendix of [3]).

The derivation of the strengthened form of the first law of black hole
mechanics is based upon the following three properties of the ADM Hamiltonian:
(i) It vanishes identically on the constraint submanifold. Hence, its first
order variation off of a
solution vanishes whenever the varied initial data satisfies the
linearized constraints. (ii) Its variation yields the Einstein-Maxwell
equations. (iii) For suitable choices of time evolution vector field in an
asymptotically flat spacetime, its variation
is directly related to formulas for the variation of mass, charge,
and angular momentum in the spacetime.

The first of these properties is manifest from eq. (6).  The second property
is just the statement of what we mean by $H$    being a Hamiltonian for
Einstein-Maxwell theory.  More explicitly, let ($h_{ab}, \pi^{ab}, A_a,
E^a$)
be initial data (satisfying the
constaints) for an Einstein-Maxwell solution, and let ($\delta h_{ab},
\delta \pi^{ab}, \delta A_a, \delta E^a$) be an arbitrary perturbation
(not necessarily
satisfying the linearized constraints) of compact support on $\Sigma$. By
integrating by parts, we can express the variation of $H$ in the
form.
\begin{equation}
\delta H = \int_{\Sigma} [ P^{ab} \delta h _{ab} + {\cal Q}_{ab} \delta
\pi^{ab} + R^a \delta A_a + S_a \delta (\sqrt{h} E^a)]
\end{equation}   
Then, the coefficient, $P^{ab}$, of $\delta h_{ab}$ yields minus the
"time derivative"
(i.e., the Lie derivative with respect to $N^\mu$) of the canonical
momentum $\frac{1}{16 \pi} \pi^{ab}$ in the solution to the Einstein-Maxwell
equations arising from the initial data ($h_{ab}, \pi^{ab}, A_a, E^a$).
Similarly, the coefficient, $16 \pi {\cal Q}_{ab}$, of $\frac{1}{16 \pi}
\delta \pi^{ab}$ yields the time derivative of $h_{ab}$, etc.
(Explicit formulas for $P^{ab}, {\cal Q}_{ab}, R^a$, and $S_a$ are given in
eqs.
(19)-(23) of [2]
in the more general case of Einstein-Yang-Mills theory.)

The third property can be understood and derived from the following
considerations [4]: Let ($h_{ab}, \pi^{ab}, A_a, E^a$) be initial data
(satisfying the constaints) for an asymptotically flat spacetime. Suppose,
now, that we consider variations ($\delta h_{ab}, \delta \pi^{ab}, \delta
A_a, \delta E^a$)
of this initial data which are merely asymptotically flat (rather than
being of compact support).  Then extra terms will appear in eq. (7) due
to contributions from boundary terms at infinity which arise when one does
the  integrations by parts needed to put the volume terms in the form
(7).  In the case where $N^\mu$ asymptotically approaches a time
transition (i.e., the lapse function $N$ goes to 1 and the shift
vector $N^a$ goes to 0 at infinity), one obtains (assuming that no
other boundaries are present on $\Sigma$ -- see below),
$$
\delta H = \int_{\Sigma} [P^{ab} \delta h_{ab} + {\cal Q}_{ab} \delta
\pi^{ab} + R^a \delta A_a + S_a \delta (\sqrt{h} E^a)]
$$
\begin{equation}
- \frac{1}{16 \pi} \delta \oint_{\infty} dS^a[\partial^b h_{ab} -
\partial_a h_b^b] - \frac{1}{4 \pi}\delta \oint _{\infty} dS^a A_0 E_a
\end{equation}   
Equation (8) suggests that we modify the definition of the ADM Hamiltonian
by addition of a surface term,
\begin{equation}
\tilde{H} \equiv H + \frac{1}{16 \pi} \oint _{\infty} dS^a [\partial^b
h_{ab} - \partial_a h_b^b] + \frac{1}{4 \pi} \oint _{\infty} dS^a A_0
E_a
\end{equation}    
If we do so, then $\tilde{H}$   will act as a true Hamiltonian on phase space
in
 the
sense that its variation will be given by the right side of eq.(7) for all
asymptotically flat perturbations. It is natural, then, to define the
``canonical energy'' $E$  on the constraint
submanifold of phase space to be the numerical value of this true Hamiltonian.
Hence, we obtain,

\begin{equation}
{\cal E} = \frac{1}{16 \pi} \oint _{\infty} dS^a [\partial^b h_{ab} -
\partial_a h_b^b] + \frac{1}{4 \pi} \oint _{\infty} dS^a A_0 E_a = m +
VQ
\end{equation}  
where $m$ is the ADM mass, $V$ is the asymptotic value of  $A_0$   at
infinity, and $Q$  is the electric charge.  (In the case presently
considered -- where there are no ``boundaries'' aside from infinity --   $Q$
will vanish, but we keep this term in eq. (10) since
it will be nonvanishing in the more general cases considered below.)
In terms of the original ADM Hamiltonian $H$, we thereby obtain,
\begin{equation}
\delta H = \int_{\Sigma} [P^{ab} \delta H_{ab} + {\cal Q}_{ab} \delta \pi^{ab}
+ R^a \delta A_a + S_a \delta (\sqrt{h} E^a)] - \delta m - V \delta Q
\end{equation}    
which yields the desired relationship between the variation of $H$  and
the variation of ADM mass, $m$, and charge, $Q$, in the case where $N^\mu$
approaches a time translation at infinity. In a similar manner, if $N^\mu$
asymptotically approaches a rotation
at infinity, we obtain,
$$
\delta H = \int _{\Sigma} [P'^{ab} \delta h_{ab} + {\cal Q}'_{ab} \delta \pi
^{ab} + R'^a \delta A_a + S'_a \delta (\sqrt{h} E^a)] + \delta J
$$
\begin{equation}
\end{equation}    
where $J$ is the "canonical angular momentum" defined by [2] (see also [5]),
\begin{equation}
J = - \frac{1}{16 \pi} \oint _{\infty} (2 \phi_b \pi^{ab} + 4
\phi^bA_bE^a)dS_a
\end{equation}
where $\phi^a$ is an asymptotic rotational Killing field on $\Sigma$.
(In eq. (12),
I have inserted primes on the quantities $P'^{ab}$, etc. appearing in the
volume integral to alert the reader to the fact that these quantities
depend upon the choice of $N^\mu$       and,
hence are different in eqs. (11) and (12), since different choices of $N^\mu$
have been made.  The Hamiltonian functions appearing on the left sides of
these equations also, of course, are different for the same reason, but
since I prefer to use $H$      to
denote the Hamiltonian (6) for any choice of $N^\mu$, I have not
inserted a prime on $H$ in eq. (12).)  Equations (11) and (12) give
explicit expression of property (iii) stated above.

The above formulas (11) and (12) are easily generalized to the case where
$\Sigma$ is a manifold with boundary, i.e., when, in addition to having an
asymptotically flat ``end'', $\Sigma$ also possesses a regular ``interior
boundary'', $S$.  In that case, the integrations
by parts needed to put the ``volume contribution'' to the variation of  $H$
in the form (7) also give rise to surface terms from $S$. These additional
surface terms are readily computed (see [2] for their explicit form).

The strengthened form of the first law of black hole mechanics follows
directly from the above three properties of the Hamiltonian formulation of
Einstein-Maxwell theory. Let ($M,g_{\mu \nu},A_\mu$) be a solution to the
Einstein-Maxwell equations describing a
stationary-axisymmetric black hole, whose event horizon is a bifurcate
Killing horizon, with bifurcation surface $S$.  Let $t^\mu$ and $\phi^\mu$
denote the Killing fields on this spacetime which, respectively,
asymptotically approach a time translation and rotation at infinity.
We assume that a Maxwell gauge choice has been made so that $A_\mu$
is nonsingular everywhere outside the black hole and on the event
horizon, and satisfies ${\cal L}_tA_\mu = {\cal L}_\phi A_\mu = 0$.
(Note that the
assumption that we can introduce a globally well defined vector
potential restricts consideration to the case where the magnetic
charge vanishes.  However, there actually is no loss of generality in
restricting attention to this case, since the magnetic charge
always can be put to zero by means
of a duality rotation.)  Let
\begin{equation}
\chi^\mu = t^\mu + \Omega \phi^\mu
\end{equation}
denote the linear combination of $t^\mu$  and $\phi^\mu$  which vanishes
on  $S$.
(Equation (14) defines the ``angular velocity of the horizon'', $\Omega$.)
Let $\Sigma$  be an asymptotically flat hypersurface which terminates on the
bifurcation surface $S$.  Let ($h_{ab}, \pi^{ab}, A_a, E^a$)
denote the initial data which is induced on  $\Sigma$.  Finally, let $H$
denote the ADM Hamiltonian asssociated with the time evolution vector field
$N^\mu = \chi^\mu$.

Now, let ($\delta h_{ab}, \delta \pi^{ab}, \delta A_a, \delta E^a$) denote
any perturbation
of the above initial data which is asymptotically flat, is nonsingular on
$\Sigma$ (including $S$), and which satisfies the linearized constraint
equations.  Then, by property (i) above,
we have $\delta H = 0$ for this perturbation.  However, by property (ii) above,
together with the fact that $\chi^\mu$  is a symmetry of the background
solution

(i.e., ${\cal L}_\chi g_{\mu \nu} = {\cal L}_\chi A_\mu = 0$),
it follows immediately that the
``volume contribution'' to $\delta H$ vanishes.
Thus it is clear from eqs. (11) and (12) that property (iii) will
give rise to a formula relating the variations in mass, angular
momentum, and charge associated with the perturbation (i.e., the
``surface terms'' from infinity) to a surface contribution from $S$
(which was not included in eqs. (11) and (12) above).  Since $N^\mu =
\chi^\mu$ vanishes on $S$, the evaluation of this boundary term from
$S$ simplifies considerably.  The final result thereby obtained is
the following [2]:  For any nonsingular, asymptotically flat
perturbation of a stationary, axisymmetric black hole with bifurcate
horizon, we have,
\begin{equation}
\delta m + V \delta Q - \Omega \delta J = \frac{1}{8 \pi} \kappa \delta A
\end{equation}  
Here, $\kappa$  denotes the surface gravity (see, e.g., [6]) of the horizon and

$A$ denotes the area of $S$. (The term $\frac{1}{8 \pi} \kappa \delta A$ is, of
course,
just the
surface contribution from $S$.)  Equation (15) expresses the first law of
black hole mechanics.

The above derivation of eq. (15) is considerably simpler than the original
derivation given in [7].  More significantly, the result obtained here is
considerably stronger: The derivation of [7] establishes that eq. (15) holds
only for perturbations to
other stationary, axisymmetric black holes.  (An extension of the derivation
of [7] to include a somewhat more general class of perturbations which are
``$t = \phi$--symmetric'' was given in [8].)  The above derivation proves that
eq. (15) holds for all nonsingular asymptotically flat perturbations
which satisfy the linearized constraint equations.  As we now shall
show, this strengthened form of the first law will enable us to close
a gap that had existed for many years in the proof of the black hole
uniqueness theorems.

\section
{Application to the Black Hole Uniqueness Theorems}
\indent
The conclusion that the charged Kerr solutions are the only stationary
black hole solutions in Einstein-Maxwell theory rests on the combined work
of many authors.  One of the key steps in the argument leading to this
conclusion is a theorem of Hawking
[9], [10], which usually is quoted as asserting that a stationary black
hole must either be static or axisymmetric.  Under the assumption that the
surface gravity, $\kappa$, is nonvanishing (corresponding to the case of a
bifurcate Killing horizon -- see [11]), Israel's theorem [12], [13]
then proves that the only static black holes in Einstein-Maxwell
theory are the Reissner-Nordstrom solutions (i.e., the charged Kerr
solutions with vanishing angular momentum), whereas the combined work
of Carter [14], Robinson [15], Mazur [16], and Bunting (see Carter
[17]) establishes uniqueness of the charged Kerr solutions in the
stationary, axisymmetric case.

However, Hawking's theorem actually states the following: First, the
theorem asserts that the event horizon of a stationary black hole must
be a Killing horizon, i.e., there must exist a Killing field $\chi^\mu$
in the
spacetime which is normal to the horizon.
If $\chi^\mu$ fails to coincide with the stationary Killing field
$t^\mu$, then
it is shown that the spacetime must be axisymmetric as well as stationary.
In that case it follows immediately that eq. (14) will hold with
$\Omega \neq 0$ -- i.e., the black hole will be
``rotating'' --and $t^\mu$ will be spacelike in a neighborhood of the horizon,
so
that the black hole will be enclosed by an ``ergoregion.''  On the other hand,
if $t^\mu$ coincides with $\chi^\mu$ (so that the black hole is ``non-
rotating'') AND if
$t^\mu$ is globally timelike outside of the black hole (so that no
``ergoregions'' exist), then it is shown that the spacetime must be
static.  However, the case where $t^\mu$ coincides with $\chi^\mu$ but
fails to be globally timelike outside of the black hole is not ruled
out by the theorem, although plausibility arguments against this possibility
have been given [10]; see also [18].
Consequently, the standard black hole uniqueness theorems leave open the
following loophole: In principle, there could exist additional stationary
black hole
solutions to the Einstein-Maxwell equations with bifurcate horizon which
are neither static nor axisymmetric. Such black holes would have to be
nonrotating (in the sense that $t^\mu$  coincides with $\chi^\mu$)
and also would
have to have a nontrivial ergoregion.  Furthermore, since $\chi^\mu$
automatically is timelike in a neighborhood of the horizon outside of
the black hole (see [19]), this ergoregion would have to be disjoint from
the horizon.  I now shall show how this loophole can be closed by
proving that any nonrotating black hole in Einstein-Maxwell
theory whose ergoregion is disjoint from the horizon must be static--even if
$t^
\mu$
is not initially assumed to be globally timelike outside of the black hole.
  In particular, this gives a direct proof that there cannot exist black
holes in Einstein-Maxwell theory whose ergoregion is disjoint from the
horizon.  The proof relies
directly upon the strengthened form of first law of black hole mechanics
obtained in the previous section, and thus provides an excellent example of
the utility of this result.

 Although the derivation of eq. (15) was given above for the case of a
stationary, axisymmetric black hole, it is immediately clear that the
derivation also applies for a black hole which is merely stationary
(i.e., which possesses a Killing field $t^\mu$
which approaches a time translation near infinity) but is non-rotating in
the sense that $t^\mu$  vanishes on $S$.  In that case, we obtain,
\begin{equation}
\delta m + V \delta Q = \frac{1}{8 \pi} \kappa \delta A
\end{equation}  
i.e., eq. (15) holds with $\Omega = 0$.  Hence, as an immediate corollary of
our strengthened form of the first law of black hole mechanics, we obtain
the following result: {\em For an arbitrary stationary, nonrotating
Einstein-Maxwell black hole, any nonsingular,
asymptotically flat perturbation of the initial data which satisfies the
linearized constraint equations, preserves the charge, {\bf Q},
of the black
hole and preserves the area, {\bf A}, of {\bf S}
cannot result in a first order
change the ADM mass, {\bf m}, of the spacetime.}

We now shall attempt to explicitly construct a perturbation which violates
this corollary.  As we shall see, this attempt will succeed unless the
spacetime is static.  Consequently, we shall conclude that every stationary,
non-rotating Einstein-Maxwell
black hole must be static.

The first (and, technically, most difficult) step in the argument is to
prove that in the (unperturbed) stationary black hole spacetime, a maximal
(i.e., vanishing trace extrinsic curvature) slice, $\Sigma$, always can be
chosen which intersects the bifurcation
surface, $S$, is asymptotically flat and is asymptotically orthogonal
to $t^\mu$ at infinity.  A proof that such a slice exists is given in [20],
and we refer the reader to that reference for further details.

Now, let ($h_{ab}, \pi^{ab}, A_a, E^a$) be the initial data which is
induced on the maximal slice, $\Sigma$, of the previous paragraph for an
(unperturbed) nonrotating Einstein-Maxwell black hole.  Consider the
following perturbation of this initial data:
\begin{equation}
\delta h_{ab} = 4 \phi h_{ab}
\end{equation}   
\begin{equation}
\delta \pi^{ab} = - 4 \phi \pi^{ab} - \pi^{ab}
\end{equation}   
\begin{equation}
\delta A_a = - A_a
\end{equation}  
\begin{equation}
\delta E^a = - 6 \phi E^a
\end{equation}   
where $\phi$  is the solution to
\begin{equation}
D^aD_a \phi - \mu \phi = \rho
\end{equation}  
on $\Sigma$  determined by the boundary conditions $\phi \rightarrow 0$
at infinity
and $\phi = 0$ on $S$, where
\begin{equation}
\mu = \frac{1}{h} \pi^{ab} \pi_{ab} + E^aE_a +
\frac{1}{2} F_{ab}F^{ab}
\end{equation}
\begin{equation}
\rho = \frac{1}{4} [\frac{1}{h} \pi^{ab} \pi_{ab} + F_{ab} F^{ab}]
\end{equation}  
Then it may be verified directly that this perturbation satisfies the
linearized constraint equations and also satisfies
$\delta Q = \delta A = 0$.  However, it also can be proven [2] that this
perturbation satisfies  $\delta m < 0$ unless $\rho =  0$.  Consequently, a
contradiction with the first law of black hole mechanics will be obtained
unless  $\pi^{ab} = 0$ (and also $F_{ab} = 0$), i.e., the first law implies
that the
full extrinsic curvature of $\Sigma$  must vanish.  By isometry invariance,
the one-parameter family of slices, $\Sigma_t$, obtained by ``time
translating'' $\Sigma$ along the orbits of $t^\mu$ also must have
vanishing extrinsic curvature.  However, it then follows that the
projection of $t^\mu$ normal to these hypersurfaces (i.e., $t^{\prime \mu} =
- (t^\nu n_\nu)n^\mu$, where $n^\mu$ is the unit normal field to
$\Sigma_t$) must be a Killing field.  (Indeed, since $t^{' \mu}$
approaches $t^\mu$ at infinity, we actually must have $t^{'\mu} =
t^\mu$.)  Hence, the nonrotating black hole possesses a hypersurface
orthogonal, timelike Killing
field which is everywhere timelike
outside the black hole and approaches a time translation at infinity.
Thus, the black hole is static, as we desired to show.

Interestingly, this proof does not generalize to Einstein-Yang-Mills case.
Indeed, it is argued in [2] that nonrotating black holes which fail to be
static will occur in Einstein-Yang-Mills theory, although such solutions,
if they exist, should be unstable.

\bibliographystyle{plain}

\end{document}